\documentclass[aps,article,twocolumn,superscriptaddress]{revtex4-2}
\usepackage{graphics} 
\usepackage{mathtools}
\usepackage{amsmath,amssymb}
\usepackage{hyperref}  
\usepackage[utf8]{inputenc}

\usepackage{graphicx}
\usepackage{color}

\begin{document}

\title{Canonical Vorticity Perspective on Magnetogenesis: Unifying Weibel, Biermann, and Beyond}

\author{Modhuchandra Laishram}\email{modhuchandra.laishram@apctp.org}
\affiliation{Asia Pacific Center for Theoretical Physics, Pohang, Gyeongbuk 37673, Republic of Korea}


\author{Young Dae Yoon}
\affiliation{Asia Pacific Center for Theoretical Physics, Pohang, Gyeongbuk 37673, Republic of Korea}
\affiliation{Department of Physics, Pohang University of Science and Technology, Pohang, Gyeongbuk 37673,
Republic of Korea}

\date{\today}

\begin{abstract}
We briefly review the current status of magnetogenesis, a cross-disciplinary field that bridges cosmology and plasma physics, studying the origin of magnetic fields in the universe. 
We formulate a canonical vorticity framework to investigate 
kinetic plasma physics-based magnetogenesis processes in a collisionless plasma. By considering canonical vorticity, a weighted sum of the fluid vorticity and the magnetic field as the canonical variable, this framework unifies several magnetogenesis processes, including the Biermann battery, the Weibel instability, and predicts several new pressure tensorial configurations as the fundamental source of self-generated magnetic field and vorticity in plasma. The framework is further extended to relativistic regime where an additional source of canonical vorticity, termed as kineclinicity effect, is identified. The theoretical predictions are systematically validated using particle-in-cell simulations, highlighting their implications for laboratory and astrophysical plasma environments.
\end{abstract}

\maketitle
The dynamical behavior of plasmas spans a wide range of regimes, from non-relativistic laboratory plasmas to relativistic and ultra-relativistic astrophysical pair plasmas~\cite{Takabe2021,Chen2023,Anglada2018,Blandford2019,Joshi2023}. Magnetic fields are ubiquitous across these systems, exhibiting an enormous range of spatial scales and strengths, from $\mu G$ fields in the interstellar medium and galaxies, to $G$-level fields around planets, stars, and black holes, and reaching up to $10^{10}G$ near neutron stars~\cite{Parker1979,Bonafede2010,Brandenburg2005,Vachaspati2021}.

Magnetic pressure plays a fundamental role in plasma dynamics by competing with other forces. For example, recent inertial confinement fusion (ICF) experiments demonstrated that magnetizing a cryogenic deuterium–tritium (DT) target with a 26-Tesla field significantly enhanced implosion performance, increasing the hot-spot temperature by 1.1 keV and the neutron yield by a factor of 2.9~\cite{Sio2023}. Magnetic fields are also essential in laboratory studies of key plasma processes such as magnetic reconnection, turbulence, collisionless shocks, and particle acceleration~\cite{Takabe2021,Remington2006,Revet2021,Bose2024}.

In astrophysical and space plasmas, magnetic fields govern a wide variety of phenomena. Planetary magnetospheres, including that of Earth, shield atmospheres from solar wind and cosmic radiation and influence auroral activity~\cite{Brandenburg2005}. In the solar system, magnetic fields drive solar flares, solar winds, and coronal heating~\cite{Guo2025,Choudhuri2007,Svalgaard2013}. On larger scales, they play crucial roles in star and galaxy formation~\cite{Vogelsberger2020,Li2018}, the acceleration and propagation of cosmic rays~\cite{Bertone2002,Andrii2010}, space weather~\cite{Griebmeier2015,Pakhotin2021}, and the overall evolution of the Universe~\cite{Turner1988}.
Despite their ubiquity and profound influence across laboratory and astrophysical plasmas, the origin of cosmic magnetic fields remains an open problem.

 From a macroscopic perspective, the evolution of magnetic fields in a plasma is governed by the ideal magnetohydrodynamic (MHD) induction equation,
\begin{eqnarray}
\frac {\partial {\bf B}}{\partial t} = \nabla\times \left({\bf u}\times {\bf B} \right)
\label{MHD_induction_eqn} 
\end{eqnarray}

Where ${\bf B}$ is the magnetic field and ${\bf u}$ the plasma mean velocity. This equation implies the frozen-flux condition: if the magnetic field is initially zero, it remains zero at all later times. In ideal MHD, therefore, the plasma provides neither a source nor a sink for magnetic fields. This fundamental limitation indicates that magnetic field generation must arise from physics beyond the macroscopic MHD description, where kinetic or non-ideal effects become important.

Most astrophysical plasmas are rotating and conducting, enabling 
convective and turbulent dynamo processes that amplify pre-existing magnetic fields, as described by Eq.~(\ref{MHD_induction_eqn}) with an independent source of ${\bf B}$~\cite{Brandenburg2012,Brandenburg2005,Mason2011}.  
It is thus widely accepted that the magnetic fields observed in various plasma systems result from the amplification of much weaker seed fields through turbulent dynamo action~\cite{Brandenburg2012}. However, the origin, scale, and structure of these initial seed fields vary significantly across different plasma environments, and remain a fundamental but poorly understood problem. 

Broadly, two main classes of theories have been proposed to explain the generation of magnetic seed fields in the Universe. The first class invokes cosmological mechanisms, in which seed fields are produced in the early Universe, such as during inflation or cosmological phase transitions prior to recombination~\cite{Kandus2011,Brandenburg2012,Andrii2010}. These models suggest that primordial magnetic fields of extremely weak strength but large coherence length can arise from symmetry breaking and quantum fluctuations, although their detailed properties remain under active debate~\cite{Vachaspati2021,Jedamzik2020}. 

And the second alternative explanations are from plasma physics aspects, which believe the seed fields are generated locally through differential plasma motions, kinetic instabilities, or non-ideal effects~\cite{Brandenburg2012,Choudhuri2007}. Specifically, in typical plasmas composed of particles with different masses, charges, pressures, and flow velocities, internal free energy can drive self-generated electric currents that act as sources of weak magnetic fields. Well-established mechanisms include the Biermann battery effect~\cite{Biermann1950}, the kinetic Weibel instability~\cite{Weibel1959}, and the electron-scale Kelvin–Helmholtz instability (ESKHI)~\cite{Gruzinov2008,Alves2012,Grismayer2013}.

The Biermann battery arises from a misalignment between electron density and temperature gradients, producing rotational electron flows that generate magnetic fields~\cite{Biermann1950,Kulsrud1997}. In contrast, the Weibel instability is a purely kinetic mechanism driven by pressure anisotropy or counter-streaming particle distributions, which are ubiquitous in most plasmas~\cite{Weibel1959,Grassi2017}. While the Biermann effect typically produces weak magnetic fields over large spatial scales~\cite{Matteucci2018,Gnedin2000}, the Weibel instability can generate relatively strong but localized magnetic fields~\cite{Huntington2015}. Similarly, the ESKHI operates in strongly sheared, typically non-relativistic flows and is often suppressed or overtaken by faster-growing Weibel modes~\cite{Liang2013,Nishikawa2014}.

The physical validity of these seed-field generation mechanisms has been supported by both laboratory experiments and astrophysical observations~\cite{Gregori2015,Zhao2024}. With the advent of high-power laser facilities, astrophysical plasma environments can now be reproduced at laboratory scales, enabling controlled investigations of magnetogenesis-relevant processes~\cite{Takabe2021,Remington2006,Revet2021}. Various laser-produced plasma experiments have confirmed Biermann battery–dominated magnetic field generation~\cite{Nilson2006,Matteucci2018} as well as Weibel-dominated magnetic field formation~\cite{Kugland2012,Huntington2015}, supporting their proposed roles in astrophysical magnetogenesis~\cite{Gnedin2000,Pucci2021,Zhou2024}. In parallel, rapid progress in astrophysical observations has provided increasing evidence consistent with theoretical predictions~\cite{Birkinshaw1996,Hardcastle2002,Widrow2002}. Measurements of synchrotron radiation, Zeeman splitting, and Faraday rotation measures of cosmic sources across large-scale structures further indicate the presence of magnetic fields with considerable coherence and strength, lending support to primordial and cosmological magnetogenesis theories~\cite{Carretti2025,Kandus2011}.

Beyond these mechanisms, there are advanced theoretical proposals for additional seed-field generation mechanisms involving relativistic effects, such as spacetime curvature-induced baroclinicity within relativistic ideal MHD~\cite{Mahajan2010,Kawazura2014}. Emerging ideas also explore magnetic field generation associated with gravitational waves or dark matter interactions~\cite{Balaji2024,Seplveda2024}. Despite these advances, the kinetic mechanisms responsible for magnetic field generation in extreme relativistic plasmas remain poorly understood and constitute an active area of research.

Consequently, research dedicated to magnetogenesis processes is growing 
driven not only by fundamental questions in plasma astrophysics and cosmology, but also by practical applications involving intense magnetic fields, including laboratory astrophysics~\cite{Remington2006,Bose2024}, fusion confinements~\cite{Sio2023,Wang2023}, and emerging technologies~\cite{Ma2022,Zhang2022}. Several relevant review papers and literature are already available on the 
topic~\cite{Savin2012,Gregori2015,Morita2023,Zhou2024}, still remain as open problems. 

The present work focuses on generalized kinetic plasma physics aspects of magnetogenesis, including relativistic effects. At the particle level, plasma species do not distinguish between inertial vorticity and the self-consistent magnetic field effects. Instead, particles respond to canonical 
momentum, which is a fundamental invariant quantity in 
Hamiltonian phase-space. At the fluid level, the corresponding canonical 
momentum become $\mathbf{P}_\sigma=m_\sigma \mathbf{u}_\sigma +q_\sigma \mathbf{A}$ whose curl is canonical vorticity ${\bf Q}_\sigma=\nabla\times \mathbf{P}_\sigma$, where $\mathbf{A}$ is magnetic vector potential, $m_\sigma$, ${q_\sigma}$, and ${\bf u}_\sigma$ are the mass, charge, and fluid velocity of the $\sigma-$species($\sigma=i,e$ for ions/electrons).
Thus, ${\bf Q}_\sigma$ rather than vorticity ${\bf \Omega}_\sigma=\nabla\times \mathbf{u}_\sigma$ or ${\bf B}=\nabla\times {\bf A}$ is the fundamental invariant quantity which dictates dynamics in the phase-space. 
Moreover, Hamilton’s equations are naturally formulated in canonical phase space $({\bf x}_\sigma, {\bf p}_\sigma)$, not $({\bf x}_\sigma, {\bf v}_\sigma)$, although both are collinear in the non-relativistic limit. This distinction becomes significant for relativistic kinetic dynamics; therefore, it is essential to track ${\bf Q}_\sigma$-variations in canonical phase space $({\bf x}_\sigma, {\bf p}_\sigma)$ for detailed studies of any dynamical process in relativistic kinetic dynamics. 
Where ${\bf x}_\sigma$, ${\bf v}_\sigma$, and ${\bf p}_\sigma$ are the  positions, velocity, and momentum of the $\sigma-$species.

In this work, in Sec.~\ref{formulation_NR}, we adopt the canonical vorticity model derived from the distribution function $f({\bf x}_\sigma, {\bf v}_\sigma, t)$ for each species composing a collisionless plasma~\cite{Yoon2025,Laishram2024,Yoon2019a}  and show that a pressure-tensor-induced mechanism named as canonical battery effect--naturally gives rise to different forms of magnetogenesis in collisionless plasma. 
In Sec.~\ref{formulation_RR}, this framework is extended for relativistic dynamics using the distribution function $f({\bf x}_\sigma, {\bf p}_\sigma, t)$\cite{Laishram2025}, and identified an additional source for canonical vorticity, termed as ``kineclinicity", which also breaks the frozen-in condition of canonical vorticity in many relativistic dynamics. 
In Sec.~\ref{Numerical_sims}, various particle-in-cell (PIC) simulations are presented to validate that the canonical battery effect unifies and generalizes both Biermann and Weibel-dominated magnetogenesis mechanisms. Using this framework, we also derive several novel pressure-tensor configurations, such as two-dimensional localized pressure anisotropies that lead to distinct magnetogenesis pathways. The role of the kineclinicity effects in relativistic magnetogenesis is demonstrated explicitly in Sec.~\ref{relativistic_sim2D}. Finally, broader implications and possible extensions of the model in both non-relativistic and relativistic regimes are discussed in Sec.~\ref{Discussions}, followed by a summary in Sec.~\ref{summary}.

\section{Canonical vorticity framework}
\label{formulation_NR}
Let us start from the following canonical vorticity equation derived from the time evolution of the 
distribution function $f({\bf x}_\sigma, {\bf v}_\sigma, t)$ of each $\sigma$-species composing a collisionless plasma~\cite{Yoon2019a,Yoon2025,Laishram2024}, 
\begin{eqnarray}
\frac {\partial {\bf Q}_\sigma}{\partial t} = \underbrace{\nabla\times \left({\bf u}_\sigma\times {\bf Q}_\sigma \right)}_{\vec{\mathcal{C}}}
\overbrace{-\nabla\times\left(\frac{\nabla\cdot \tensor{\mathcal{P}}_\sigma}{n_\sigma}\right)}^{\vec{\mathcal{B}}}.
\label{canonical_vorticity_eqn} 
\end{eqnarray}
Where $n_\sigma=\int f d^3{\bf v}$ is density, ${\bf u}_\sigma=\frac{1}{n_\sigma} \int   {\bf v}_\sigma f d^3{\bf v}$ is mean velocity, and $ \tensor{\mathcal{P}}_\sigma = m_\sigma\int {\bf v'}_\sigma{\bf v'}_\sigma f({\bf v}_\sigma)d^3{\bf v}_\sigma $ is the pressure tensor when $\mathbf{v}'_\sigma =\mathbf{v}_\sigma-\mathbf{u}_\sigma$ is the random part of $\mathbf{v}_\sigma$.  
${\bf Q}_\sigma={m_\sigma{\bf\Omega}_\sigma+q_\sigma{\bf B}} $ is the canonical vorticity, which represents the weighted sum of inertial ${\bf\Omega}_\sigma$ 
and ${\bf B}$ effects. Eqn.~(\ref{canonical_vorticity_eqn}) 
describes the time evolution of $\mathbf{Q}_\sigma$ with just two terms on the right-hand side. 
The first term is the inertial flux-conserved convective term $\vec{\mathcal{C}}$, which signifies that $\mathbf{Q}_\sigma$ is frozen and evolves with plasma's $\mathbf{u}_\sigma$. The second term is 
the canonical battery term $\vec{\mathcal{B}}$, which acts as a source/sink of $\mathbf{Q}_\sigma$ due to the kinetic pressure tensor effect. 

Following the isomorphism with the magnetic induction term in ideal MHD (Eq.~\ref{MHD_induction_eqn}) with no pressure tensor 
effect, Eq.~(\ref{canonical_vorticity_eqn}) becomes,
\begin{eqnarray}
\frac {\partial {\bf Q}_\sigma}{\partial t} = \underbrace{\nabla\times \left({\bf u}_\sigma\times {\bf Q}_\sigma \right)}_{\vec{\mathcal{C}}}.
\label{canonical_vorticity_eqn1} 
\end{eqnarray}
 
 It simply means $\mathbf{Q}_\sigma$ is frozen with $\mathbf{u}_\sigma$ and amplified along with the flow. 
 In another words, $\mathbf{Q}_e$-flux is a conserved quantity in kinetic scales 
 while $\mathbf{B}$-flux is conserved only in the ideal MHD limit. As a consequence, the flux associated with $\mathbf{Q}_\sigma$ becomes a conserved quantity in the reference frame moving with $\mathbf{u}_\sigma$. This $\vec{\mathcal{C}}$-term is important for dealing with flow characteristics related to an initially strong mean 
 flow $\mathbf{u}_\sigma$ (with $\mathbf{\Omega}_\sigma \ne 0$) such as electron scale Kelvin-Helmholtz instability(ESKHI)~\cite{Gruzinov2008, Alves2012} which 
 leads to a new form of magnetogenesis by 
 transforming shear free-energy into magnetic energy.
 
The second term $\vec{\mathcal{B}}$ is the only non-ideal source/sink that can violate the frozen-in flux 
condition of $\mathbf{Q}_\sigma$. Note that there are two ways to violate magnetic flux conservation~\cite{Yoon2025PoP}. One of the ways is magnetic reconnection, in which there is initial finite magnetic flux, and it decreases along with the flow evolution, which is common in many 
strongly magnetized plasmas~\cite{Bulanov1992,Cafaro1998,Grasso2001}.  
Another way is magnetogenesis, in which magnetic flux starts from zero or a finite value and increases along with the 
temporal evolution of the flow. Magnetogenesis is the primary focus of the present work; therefore, the following 
analysis examines how the non-ideal $\vec{\mathcal{B}}$ term is responsible for the magnetogenesis processes.

Again, it is clear from Eq.~(\ref{canonical_vorticity_eqn}) that if the plasma is initially unmagnetized ($\mathbf{B}=0$) and vorticity-free ($\mathbf{\Omega}_\sigma=0$) so that $\mathbf{Q}_\sigma=0$ and $\vec{\mathcal{C}}=0$, then Eq. (\ref{canonical_vorticity_eqn}) reduce to, 
\begin{eqnarray}
\frac {\partial {\bf Q}_\sigma}{\partial t} = 
\overbrace{-\nabla\times\left(\frac{\nabla\cdot \tensor{\mathcal{P}}_\sigma}{n_\sigma}\right)}^{\vec{\mathcal{B}}},
\label{canonical_vorticity_eqn2} 
\end{eqnarray}
It means the only term that can generate a finite $\mathbf{Q}_\sigma$ with such an initial 
condition is $\vec{\mathcal{B}}$. The generated $\mathbf{Q}_\sigma$ cannot be entirely comprised 
of $\mathbf{\Omega}_\sigma$ or ${\bf B}$ because Amp\`{e}re's law gives $-\nabla^2\mathbf{B}= \mu_0 \nabla\times\mathbf{J}\simeq\mu_0\sum_\sigma n_\sigma q_\sigma \mathbf{\Omega}_\sigma$, so 
finite $\mathbf{\Omega}_\sigma$ means finite $\mathbf{B}$. 

To understand how generation of $\mathbf{Q}_\sigma$ leads to self-generation of $\mathbf{B}$ and its characteristics, let us examine a simpler plasma system where the current is carried entirely by electrons and ions are assumed cold 
background medium, i.e., ${\bf J}=n_eq_e{\bf u}_e$ (electron-MHD assumption~\cite{Gordeev1994}). Then, Maxwell's equations give 
$\nabla^2\mathbf{B}= \mu_0 n_e q_e \mathbf{\Omega}_e$. Noting that $\sqrt{m_e/\mu_0 n_e e^2}=d_e$ is the 
electron skin depth, the electron canonical vorticity becomes $\mathbf{Q}_e=-q_e\left(d_e^2\nabla^2\mathbf{B}-\mathbf{B}\right)$ and is expressed entirely as a function of $\mathbf{B}$. Now, using spatial Fourier transform $\tilde{\mathbf{Q}}_e=\int\mathbf{Q}_e \exp\left(-i\mathbf{k}\cdot\mathbf{x}\right)d^3\mathbf{x}$, and 
rearrange the equation for $\mathbf{B}$, we have
\begin{equation}
    \mathbf{B}=\frac{1}{2\pi q_e}\int \frac{\tilde{\mathbf{Q}}_e}{1+k^2 d_e^2} \exp\left(i\mathbf{k}\cdot\mathbf{x}\right)d^3\mathbf{k}.
\label{B_in_terms_of_Q}
\end{equation}
Equation (\ref{B_in_terms_of_Q}) shows that for $k^2 d_e^2\gg 1$ (i.e, $d_e\gg \lambda$, small scales), $\mathbf{B}$ is small and thus $\mathbf{Q}_e\simeq m_e\mathbf{\Omega}_\sigma$.
 whereas for $k^2 d_e^2\ll 1$ (i.e, $d_e\ll \lambda$, large scales), $\mathbf{Q}_e\simeq -e\mathbf{B}$. 
 Further, for $kd_e\sim 1$, the inertial and magnetic terms contribute equally. 
 In other words, $\mathbf{B}$ is a ``smoothed out'' function of $\mathbf{Q}_e$ in kinetic scales. 
 Thus, generation of $\mathbf{Q}_e$ by the canonical battery effect 
 essentially leads to generation of $\mathbf{B}$ more in the  $k^2 d_e^2\ll 1$ scales. 
 
 Keeping the above understanding in mind, let us now explore the pressure-tensor components in 
 Eq.~(\ref{canonical_vorticity_eqn2}) for a 2D flow system assuming 
 $\partial/\partial z \rightarrow 0$, and so the $z$-component of $\vec{\mathcal{B}}$ becomes~\cite{Laishram2024},
\begin{multline}
\mathcal{B}_{z} =\frac{1}{n_e} \left(\frac{\partial^2}{\partial y^2}-\frac{\partial ^2}{\partial x^2}\right)p_{exy}
+\frac{1}{n_e}\frac{\partial^2}{\partial x \partial y}\left(p_{exx}-p_{eyy}\right)\\
+ \hat{z}\cdot\left[-\nabla\left(\frac{1}{n_e}\right)\times\nabla\cdot\tensor{\mathcal{P}}_e\right]
\label{canonical_battery_2Dz} 
\end{multline}
where $p_{eij}$ is the $ij$-component of $\tensor{\mathcal{P}}_e$. One of the significant aspects of this model is that each term on the right side of 
Eq.~(\ref{canonical_battery_2Dz}) corresponds to different kinetic processes or instabilities. In the following, we will see 
how $\mathcal{B}_{z}$ generalized 
the well-known mechanisms of magnetogenesis showing each terms of $\mathcal{B}_{z}$ is responsible for various form 
of magnetogenesis, such as the Biermann Battery effect, the Weibel Instability, and newly found additional effects solely visible through this formulation. 

\subsection{$\mathcal{B}_{z}-$ for Weibel Instability}
\label{Weibel}
The Weibel instability is a fundamental kinetic phenomenon commonly encountered in a wide range of plasmas, as its primary requirement is 
merely a directional anisotropy in the thermal motion of charged particles~\cite{Weibel1959}. As a result, it naturally arises during the isotropization process of 
any initially anisotropic thermal distribution. While this instability can be observed in any spatial dimension, it is notable that variation along a single spatial direction is sufficient to trigger and observe its development. 
For clarity, let us assume a 1D system ($\nabla \rightarrow \hat{\bf x}\partial/\partial x$) with an initially uniform 
density ($\nabla n_e =0$), then only the $1st-$term in Eq.~(\ref{canonical_battery_2Dz}) survives and becomes
\begin{align}
    \mathcal{B}_{z} =-\frac{1}{n_e} \left(\frac{\partial^2 p_{exy}}{\partial x^2}\right).
    \label{B_z_weibel}
\end{align}
Although Weibel instability is generally attributed to pressure anisotropy $p_{eyy} \ne p_{exx}$, Eq. (\ref{B_z_weibel}) explicitly shows that the off-diagonal term $p_{exy}$ arising from degree of mixing between initially unequal diagonal components $p_{exx} $ and $p_{eyy}$ is what actually generates $Q_{ez}$ in such simplified plasma systems. 
Thus, the $1st-$term of the canonical battery $\mathcal{B}_{z}$ in Eq.~(\ref{canonical_battery_2Dz}) embodies the magnetogenesis through the Weibel instability. In section~\ref{Weibel_sim1D}, we will visualize numerically how $p_{exy}$ excites the Weibel instability in 1D and 2D collisionless plasma systems and generate filamentary structure 
of magnetic fields.

\subsection{$\mathcal{B}_{z}-$ for Localized Pressure Anisotropy Effect}
\label{localizedP}
When we extend the above analysis to a 2D system 
($\nabla \rightarrow \hat{\bf x}\partial/\partial x+\hat{\bf y}\partial/\partial y$ and $\hat{\bf z}\partial/\partial z\rightarrow0$) with the same 
uniform density ($\nabla n_e =0$) and pressure anisotropy ($p_{eyy} \ne p_{exx}$), both $1st-$ and $2nd-$terms  
in Eq.~(\ref{canonical_battery_2Dz}) survives and becomes,
\begin{multline}
\mathcal{B}_{z} = \frac{1}{n_e} \left(\frac{\partial^2}{\partial y^2}-\frac{\partial ^2}{\partial x^2}\right)p_{exy}
+\frac{1}{n_e}\frac{\partial^2}{\partial x \partial y}\left(p_{exx}-p_{eyy}\right).
\label{canonical_battery_2Dz_local} 
\end{multline}
Here, the $1st-$term involves the off-diagonal component and so is responsible for the Weibel instability in a 2D system. The $2nd-$term differs from the Weibel instability effect discussed above, which is an additional local effect involving exact 2D-localized pressure anisotropy. 
For instance, if the initial localized anisotropy is of the form $p_{exx}-p_{eyy}\sim \exp\left[-(x^2+y^2)/\sigma^2\right]$ with an uniform density, then the last term in Eq.~(\ref{canonical_battery_2Dz_local}) can be dominant and has the structure as
\begin{align}
    \mathcal{B}_z \sim xy \exp\left(-\frac{x^2+y^2}{\sigma^2}\right),
    \label{2D_localized_anisotropy}
\end{align}
which predicts that a quadrupole magnetic field will be locally generated in such a 2D system. Thus, dominant terms in Eq.~(\ref{canonical_battery_2Dz}) enable us to generally predict the characteristics of the pressure tensor configurations and associated magnetogenesis processes. In section~\ref{Weibel_sim2D}, we will display numerically how such localized anisotropies indeed generate a quadrupole magnetic field in 2D collisionless plasma system and vary their characteristics.

\subsection{$\mathcal{B}_{z}-$ for Biermann Battery Effect}
\label{Biermann}
The Biermann battery effect~\cite{Biermann1950} is known for magnetogenesis in plasma having isotropic pressure and baroclinicity nature when the electron's pressure and density gradients are not aligned. So it's not about pressure anisotropy, but baroclinicity ($\nabla{n_e } \times  \nabla{p_e}\ne0$), therefore, this idea 
is theorized to be important in a wide range of scales from cosmological to laboratory magnetogenesis~\cite{Gregori2015}. In such systems, 
$\tensor{\mathcal{P}}_e  =p_e \tensor{\bf { I}}$,  where $ \tensor{\bf { I}}$ is the identity 
tensor, and only the $3rd-$term in Eq.~(\ref{canonical_battery_2Dz}) survives and becomes,
\begin{eqnarray}
\mathcal{B}_{z}
= \hat{z}\cdot\left[-\nabla\left(\frac{1}{n_e}\right)\times\nabla p_e\right]
=  \hat{z}\cdot\left[\frac{\nabla{n_e } \times  \nabla{p_e}} {n_e^2}\right].
\label{Biermann_eqn2} 
\end{eqnarray}

Here, $\mathcal{B}_{z}$ becomes finite when there is a misalignment of density and 
pressure, which is the essential condition for the 
Biermann battery 
effect~\cite{Biermann1950,Kulsrud1997,Matteucci2018}.
Note that this $3rd-$term in
Eq.~(\ref{canonical_battery_2Dz}) can be considered as a generalization of the Biermann battery effect 
because it retains the full pressure tensor representing a kinetic form of the Biermann battery. Moreover, this formulation extends the Biermann concept to all physical regimes, from the kinetic to the MHD scales. 
In Section~\ref{Beirman_sim2D}, we demonstrate numerically how this generalized formulation captures the 
kinetic-scale Biermann battery effect, even though this effect is typically associated with 
isotropic MHD-scale processes.

\section{Relativistic canonical vorticity model}
\label{formulation_RR}
When the above model resulting Eq.~(\ref{canonical_vorticity_eqn}), is reformulated for relativistic regimes in the canonical momentum space using $f({\bf x}_\sigma, \bf{p}_\sigma, t)$ instead of the $f({\bf x}_\sigma, {\bf v}_\sigma, t)$ using transformation $\mathbf{p}_\sigma=\gamma m_{\sigma} {\bf v}_\sigma $, where $\gamma$ is the Lorentz factor~\cite{Laishram2025,Melzani2013}, the equation for canonical vorticity ${\bf Q}_\sigma$ becomes,
{\small
\begin{eqnarray}
\frac {\partial {\bf Q}_\sigma}{\partial t} = \underbrace{\nabla\times \left({\bf u}_\sigma\times {\bf Q}_\sigma \right)}_{\vec{\mathcal{C}}}
\overbrace{-\nabla\times\left(\frac{\nabla\cdot \tensor{\mathcal{P}}_\sigma}{n_\sigma}\right)}^{\vec{\mathcal{B}}} 
+  \underbrace{J[\left<\mathbf{p}\right>_\sigma,{\bf u}_\sigma]}_{\vec{\mathcal{R}}}.~~~~
\label{canonical_vorticity_eqn_rel} 
\end{eqnarray}
}

Where, $n_\sigma =\int f_\sigma d^3\mathbf{p}$ is the density, $\mathbf{u}_\sigma=n_\sigma^{-1}\int \mathbf{v} f_\sigma d^3\mathbf{p}$ is the fluid velocity, $\left<\mathbf{p}\right>_\sigma=n_\sigma^{-1}\int \mathbf{p} f_\sigma d^3\mathbf{p}$ is the fluid momentum, and 
$ \tensor{\mathcal{P}}_\sigma = \int {\bf v'}{\bf p'} f_\sigma d^3{\bf p}$ is the pressure tensor, 
when $\mathbf{v}' =\mathbf{v}-\mathbf{u}_\sigma$ and $\mathbf{p}' =\mathbf{p}-\left<\mathbf{p}\right>_\sigma$
are the random part of $\mathbf{v}_\sigma$ and $\mathbf{p}_\sigma$, respectively in the canonical phase space. 

Now, compared to the above non-relativistic counterpart (Eq.~(\ref{canonical_vorticity_eqn})), 
relativistic effects add a source/sink term $\vec{\mathcal{R}}$ for the ${\bf Q}_\sigma$, 
named as ``kineclinicity-effect'', in addition to the relativistic form of canonical battery 
$\vec{\mathcal{B}}$-effects. Mathematically, the expression for $\vec{\mathcal{R}}$ of the 
$\sigma$-species in a 2D flow system ($\partial/\partial z \rightarrow 0$) turns out to be the 
sum of Jacobian determinants (or Poisson brackets) 
$\{\langle p \rangle_{j},\, u_{j}\}_{xy}$ as follows:

\begin{small}
\begin{eqnarray}
\mathcal{R}_{\sigma z}
= \sum_{j}
\left[
\frac{\partial \langle p_j \rangle_\sigma}{\partial x}\,
\frac{\partial u_{j,\sigma}}{\partial y}
-
\frac{\partial \langle p_j \rangle_\sigma}{\partial y}\,
\frac{\partial u_{j,\sigma}}{\partial x}
\right].
\label{relativistic_effect_eqn}
\end{eqnarray}
\end{small}

where $j=x,y,z$ (i.e.\ $\mathcal{R}_k=\epsilon_{ijk}\,\partial_i \langle p_\alpha\rangle\,\partial_j u_\alpha$ in index notation). 
$\mathcal{R}_{k}$ signifies the misalignment between $\langle p_\alpha\rangle$ and ${\bf u}$ and 
reduces to zero in non-relativistic regimes. The complete formulation and detailed analysis of the 
role of $\mathcal{R}_{k}$ is discussed in recent work~\cite{Laishram2025}. In 
Section~\ref{relativistic_sim2D}, we demonstrate a pair-plasma system numerically where 
$\mathcal{R}_{\sigma z}$ plays the main role in breaking the frozen-in property of $Q_{ez}$, 
giving rise to the seed $\mathbf{B}$-field.

\begin{figure*}
\includegraphics[width=\textwidth]{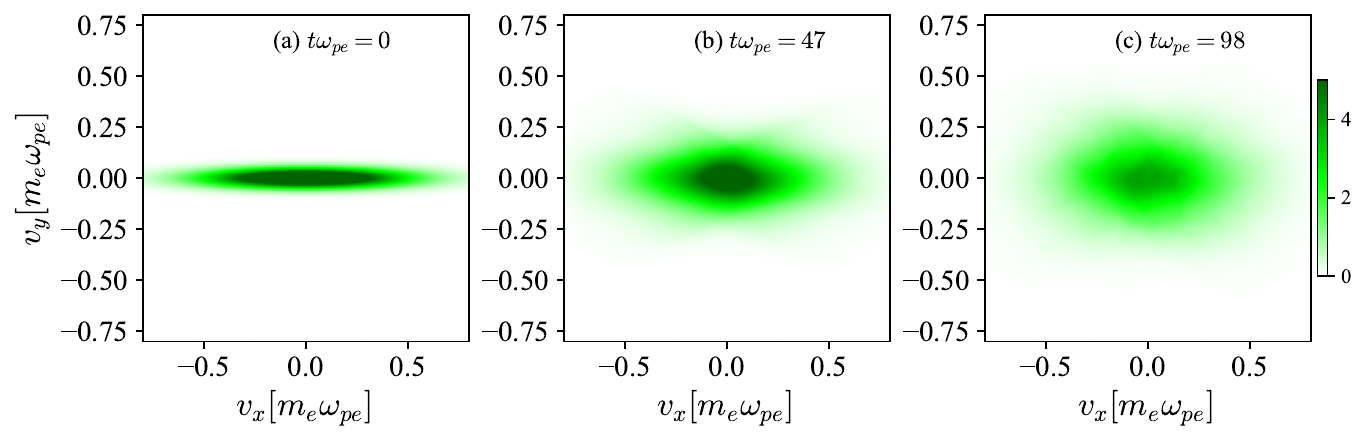}\\
\caption{Distribution of particles in $(v_x, v_y)$ space at different times (a) $t\omega_p= 0$, (b) $t\omega_p= 47$, and (c) $t\omega_p= 98$. The distribution tries to isotropize the initial pressure anisotropy as the instability proceeds~\cite{Yoon2025PoP}.}
\label{Phase_space_vx_vy}
\end{figure*}
\begin{figure*}
\includegraphics[width=0.8\textwidth]{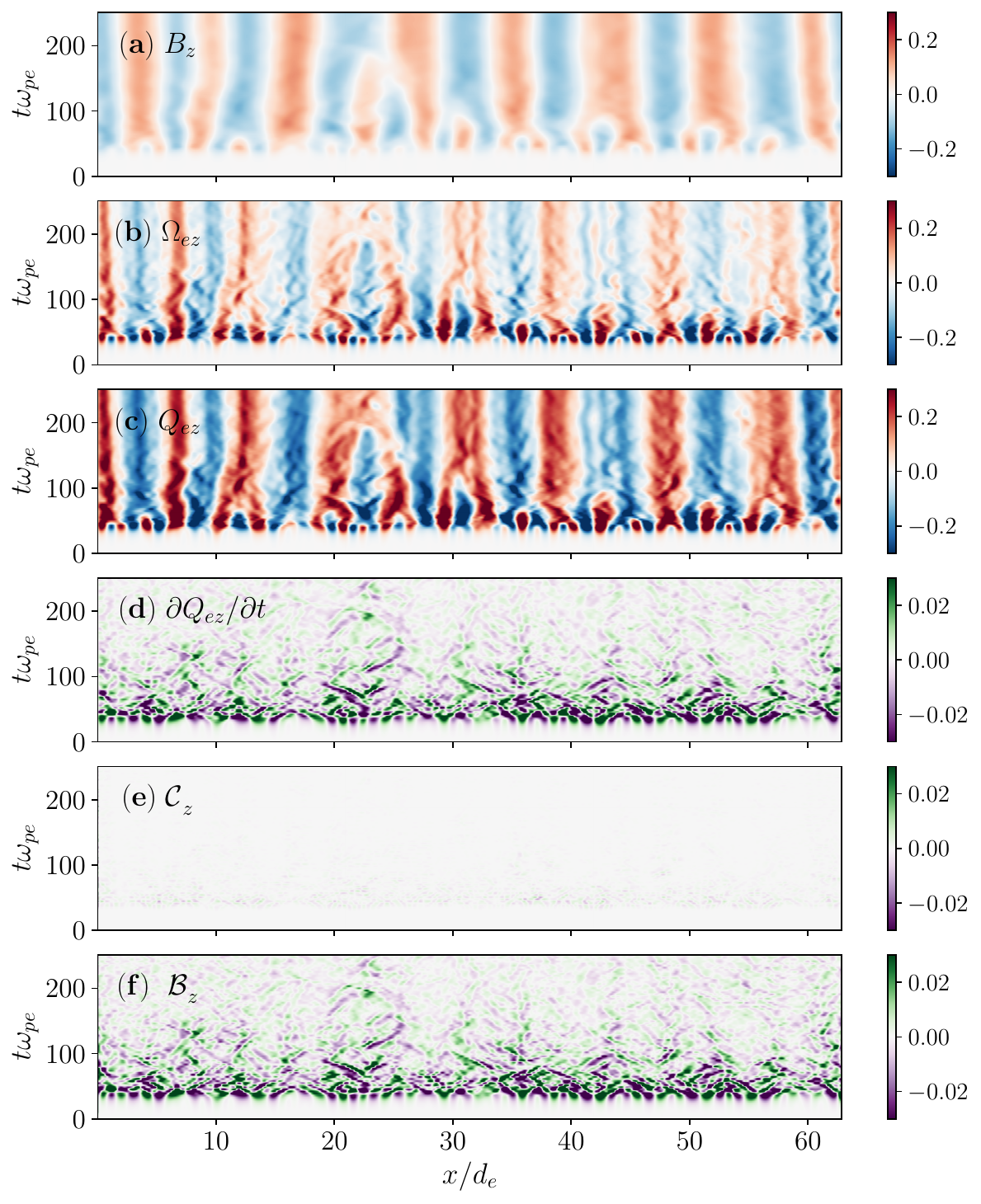}
\caption{Streak plot of various quantities of the $z$-component of
Eq.~(\ref{canonical_vorticity_eqn}), namely
(a) $B_z$,
(b) $\Omega_{ez}$, (c) $Q_{ez}$, (d) $\partial Q_{ez}/\partial t$, (e) $\mathcal{C}_z$, and (f) $\mathcal{B}_z$ from the 1D PIC simulation. They are shown in units of (a) $m_e\omega_{pe}/e$, (b) $\omega_{pe}$, (c) $m_e\omega_{pe}$, and (d-f) $m_e\omega_{pe}^2$, where $\omega_{pe}=\sqrt{n_0e^2/m_e\epsilon_0}$. }
\label{Spatiotemporal_all_By_case}
\end{figure*}

\begin{figure*}
\includegraphics[width=1.20\columnwidth]{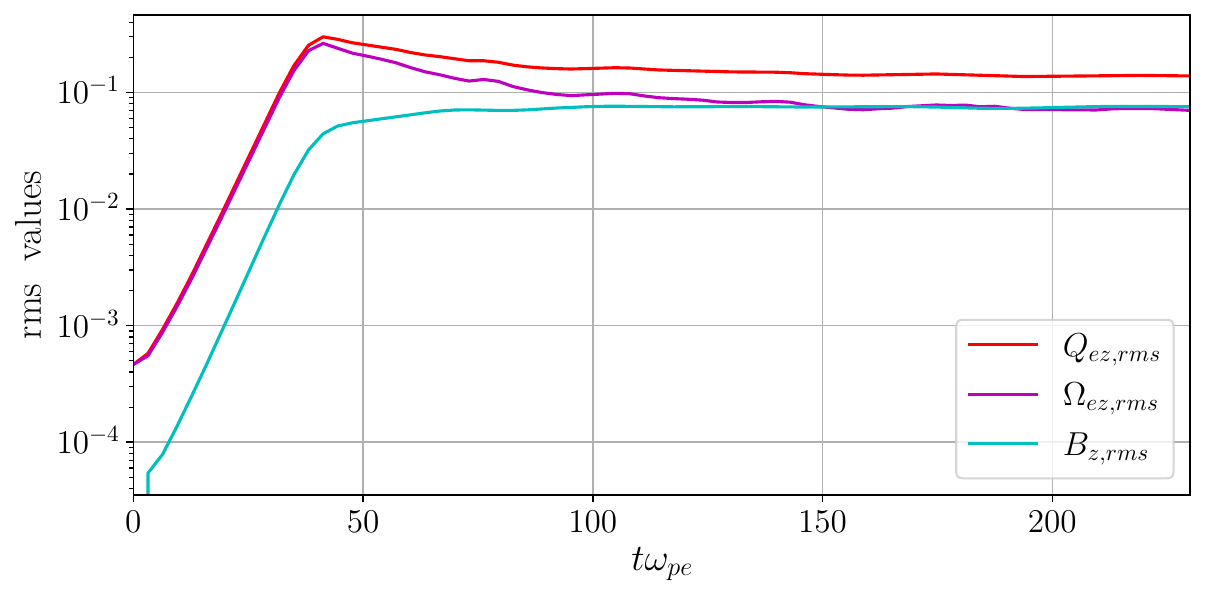}\\
\caption{Time-dependent rms values of normalized $B_z$, $\Omega_{ez}$, and $Q_{ez}$.}
\label{Spatiotemporal_all_rms_fig}
\end{figure*}
\section{Numerical Analysis}
\label{Numerical_sims}
To facilitate more understanding and visualization of the above 
theoretical predictions in Sections~\ref{formulation_NR} and ~\ref{formulation_RR}, a series of particle-in-cell(PIC) simulation were conducted using the SMILEI code~\cite{DEROUILLAT2018351} and demonstrate the different roles of canonical battery $\mathcal{B}_z$ and $\mathcal{R}_z$ in 
magnetogenesis as follows.

 \subsection{1D view of pressure anisotropy effects}
\label{Weibel_sim1D}
To validate the role of ${p}_{exy}$ predicted by Eq.~(\ref{B_z_weibel}) producing Weibel instability, several 1D3V particle-in-cell (PIC) simulation are conducted in a domain of $L_x= 20\pi d_e$ with periodic boundary conditions, divided by 2048 cells, and $10^5$ particles are 
placed per cell. The time step is $\Delta t\approx 0.0291/\omega_{pe}$ and density is set uniformly for both 
electrons and ions with a realistic ion-to-electron mass ratio $m_i/m_e=1836$. The initial ${p}_{exy}$ is generated from a 
bi-Maxwellian temperature configuration $T_{eyy}=T_{ezz}=0.1m_ec^2$ and $T_{exx}=0.001m_ec^2$, whereas ions remain as a cold background medium. 

Fig.~\ref{Phase_space_vx_vy} shows the temporal evolution of the distribution function in $(v_x, v_y)$ space, displaying  
isotropization of the initial bi-Maxwellian anisotropy to a more relaxed state through phase mixing. Note that 
${p}_{exy}$ signify the degree of the phase mixing that source the $\mathcal{B}_z$ in Eq.~(\ref{B_z_weibel}). 
Similar phase space dynamics are common in plasma systems with different initial tensorial anisotropies depending on possible combinations of $T_{eyy}$ and $T_{ezz}$ with respect to $T_{exx}$. For example, plasma with a counter-streaming of electron beams always has 
similar phase mixing dynamics~\cite{Grassi2017,Yoon2025PoP}. 
Fig.~\ref{Spatiotemporal_all_By_case} shows the corresponding streak plot of various quantities 
in Eq.~(\ref{canonical_vorticity_eqn}) and its constituent variables. All quantities have been Gaussian filtered by 10 grid 
points to reduce noise. It can be seen that $B_z$ gradually self-generates into a filamentary structure and 
saturates by $t\omega_{pe}\simeq 100$ (Fig~\ref{Spatiotemporal_all_By_case}(a)). The phase 
mixing at the electron inertial scale produces bunching of electrons, which gradually turns into current filaments and reinforces any small electromagnetic perturbation fields or existing fields.  This is indeed the fundamental characteristic of the well-known Weibel instability~\cite{Grassi2017,Weibel1959}. Further, the electron vorticity $\Omega_{ez}$ grows along with $B_z$, and together they constitute $Q_{ez}=\Omega_{ez}-B_z$ in normalized units (Figs.~\ref{Spatiotemporal_all_By_case}(b) and (c)). As expected from Eq.~(\ref{B_in_terms_of_Q}), $B_z$ has a similar profile to $Q_{ez}$ but without the finer sub-$d_e$ scale structures and with the sign of $q_e$. Differentiating $Q_{ez}$ with respect to time, one obtains $\partial Q_{ez}/\partial t$ (Fig.~\ref{Spatiotemporal_all_By_case}(d)) or the left-hand side of Eq.~(\ref{canonical_vorticity_eqn}). 
The two terms on the right-hand side are shown in Figs. \ref{Spatiotemporal_all_By_case}(e) and (f). It is evident that $\mathcal{C}_z$ is negligible and 
$\partial Q_{ez}/\partial t\approx\mathcal{B}_z$, i.e., the canonical battery effect in Eq.~\ref{B_z_weibel} is entirely responsible for the self-generation of $Q_{ez}$, which is partly $\Omega_{ez}$ and $B_z$. The growth rate of the instability can be analyzed from the temporal evolution of the root-mean-square (rms) or maximum value of $ Q_z$ and its components, as shown in Fig.~\ref{Spatiotemporal_all_rms_fig}. It displays a well-defined linear 
exponential growth phase for all quantities up to the $t\omega_{pe}\simeq 100$ with dominance of $\Omega_{ez}$, and followed by nonlinear saturation phases, 
where both $\Omega_{ez}$ and $B_z$ reach comparable magnitudes or equipartition of energies. 
Additional analysis on the evolution of the rms value and its connection to pressure tensor components 
described in Eq.~(\ref{canonical_vorticity_eqn}) are 
discussed in our recent work~\cite{Laishram2024}. However, the kinetic physics governing the nonlinear saturation phases and transition to turbulence is not straightforward~\cite{Achterberg2007, Schekochihin2009, Grassi2017} and left for future work. 

\begin{figure*}
\includegraphics[width=\textwidth]{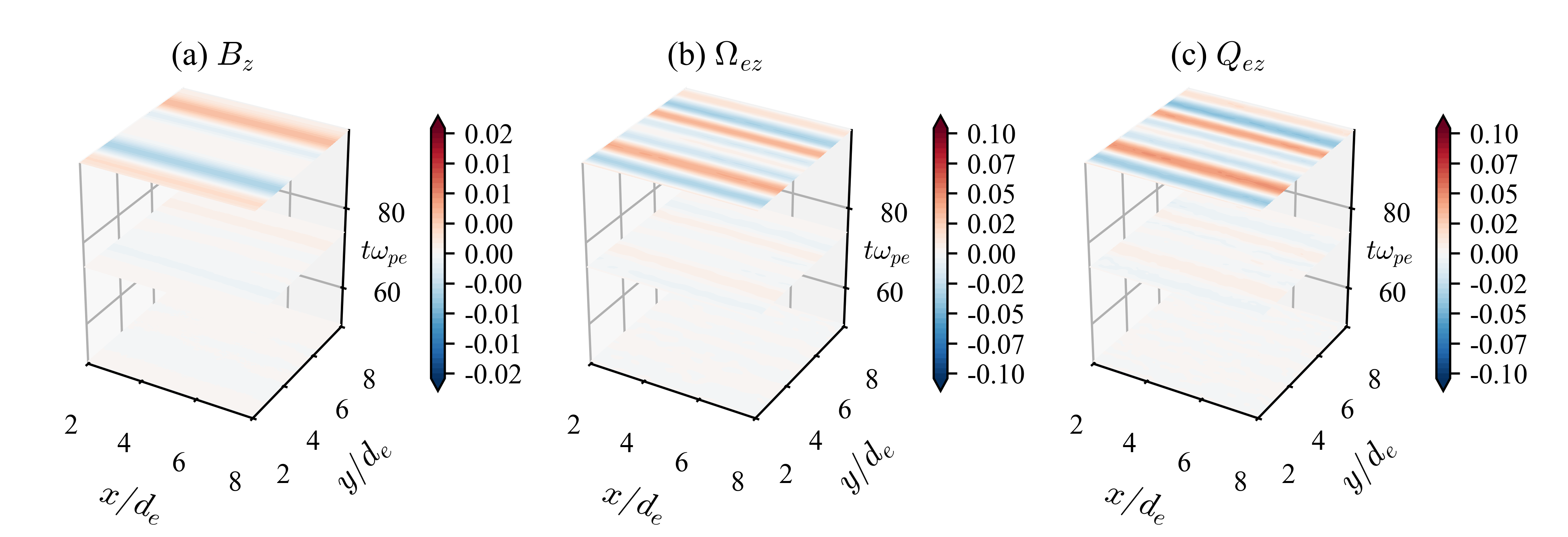}\\
\caption{Streak plots of (a) $B_z [m_e\omega_{pe}/e] $, (b) $\omega_{ez} [\omega_{pe}] $, and 
(c) $ Q_{ez} [m_e\omega_{pe}]$ with each panel corresponds to $t\omega_{pe}=33.89, 58.82,$ and $86.74$ from 
the 2D PIC simulation with uniform density and non-localized temperature anisotropy $T_{eyy}=T_{ezz}=10^{-4}m_ec^2$ and $T_{exx}=0.04 m_ec^2 $.}
\label{strek_plot_BWQ_pxy}
\end{figure*}

\subsection{2D-localized effect of pressure anisotropy}
\label{Weibel_sim2D}
To see the relative role of ${p}_{exy}$ 
and $({p}_{exx}-{p}_{eyy})$ terms in Eq.~(\ref{canonical_battery_2Dz_local}), the above simulation is extended to a 2D3V PIC simulation in a domain $(L_x,L_y)=(10,10)d_e$ divided into 1048 cells in each direction. 1000 particles were placed per cell and time-step becomes $\Delta t=6.5\times10^{-3}/\omega_{pe}$. The same initial bi-Maxwellian anisotropy with $T_{eyy}=T_{ezz}=10^{-4}m_ec^2$ 
and $T_{exx}= 10^{-2}m_ec^2$, uniform density $n_0$, cold ions background, and same periodic boundary conditions are reconsidered. 

Fig.~\ref{strek_plot_BWQ_pxy} shows streak plots of $B_z, ~\Omega_z$ and $Q_{ez}$ from the 2D simulation. All quantities 
have been Gaussian filtered by 10 grid points to enhance the visualization of large-scale structures. 
The 2D streak plots clearly illustrate how the out-of-plane magnetic field $B_z(x,y)$ self-organizes into filamentary 
structures along with time, accompanied by similar patterns in the flow vorticity $\Omega_z(x,y)$. Together, they contribute to the evolution of the canonical vorticity $Q_{ez}=\Omega_{ez}-B_z$. Since the anisotropy is non-localized in space, the contribution of the second term involving $({p}_{exx}-{p}_{eyy})$ in Eq.~(\ref{canonical_battery_2Dz_local}) is negligible. As a result, the whole dynamics is predominantly governed by the first term associated with $p_{xy}$ only. 
The spatial structures observed in $B_z(x,y)$ and $\Omega_z(x,y)$ reflect the spontaneous emergence of current seed 
filamentary structure arising during the relaxation process of the pressure anisotropy by transforming associated free energy into magnetic energy. 
\begin{figure*}
\includegraphics[width=\textwidth]{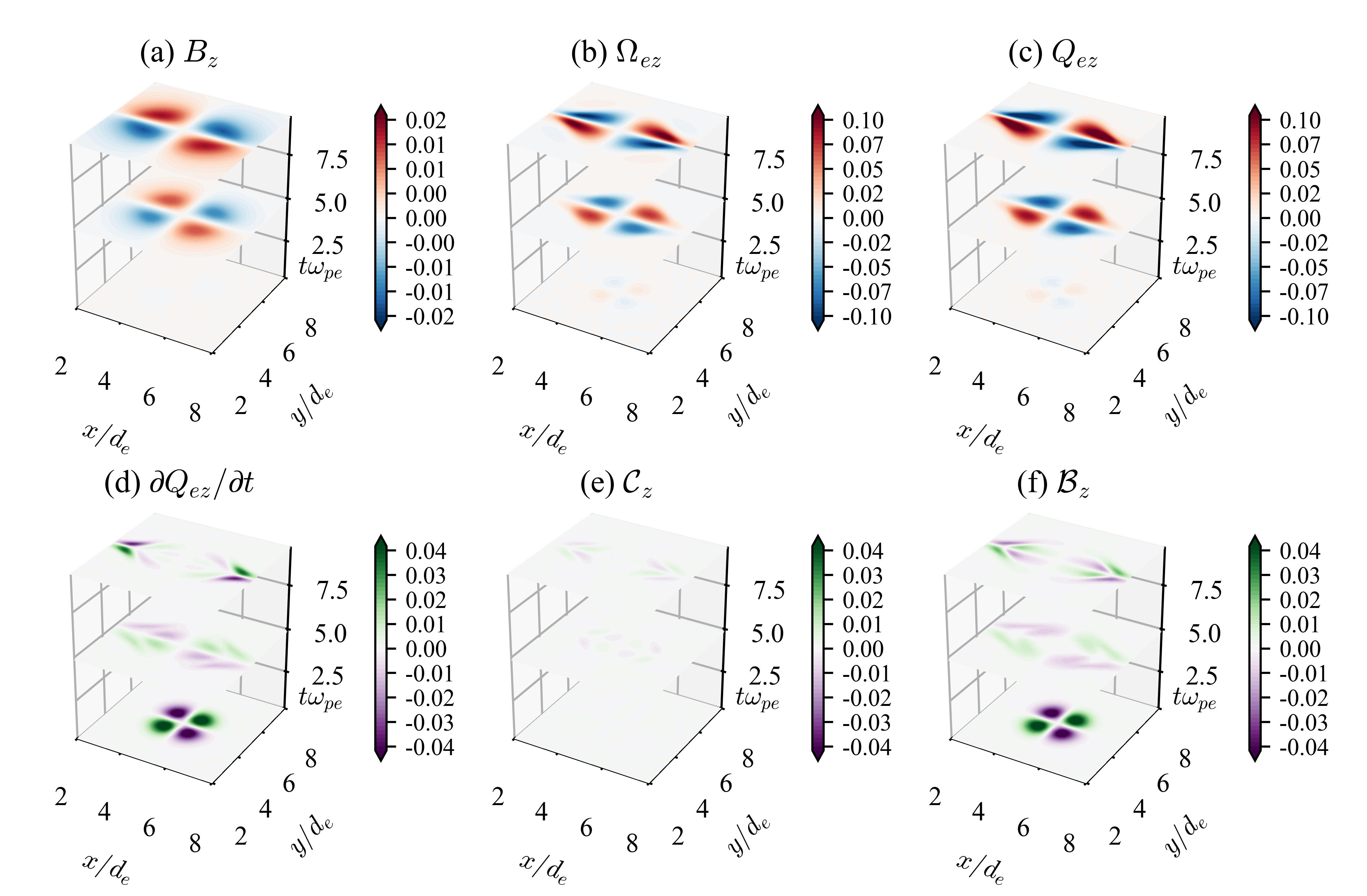}\\
\caption{Streak plots of  (a) $B_z$, (b) $\Omega_{ez}$, (c) $Q_{ez}$, (d) $\partial Q_{ez}/\partial t$, (e) $\mathcal{C}_z$, and (f) $\mathcal{B}_z$ the 2D PIC simulation 
with uniform density ($n_0$) and localized temperature 
anisotropy $T_{eyy}=T_{ezz}=10^{-4}m_ec^2$ and  
$T_{exx}=T_{eyy}+0.04m_ec^2\exp\left(-[(x-x_0)^2+(y-y_0)^2]/\delta^2\right)$. 
The units are the same as in Fig. \ref{Spatiotemporal_all_By_case}. 
The three slices in each panel correspond to $t\omega_{pe}=$ 0.19, 5.02, and 9.94.}
\label{2D_streak_xx_yy}
\end{figure*}

Now, to examine the influence of the $({p}_{exx}-{p}_{eyy})$ term in Eq.~(\ref{canonical_battery_2Dz_local}), we now modify the initial bi-Maxwellian anisotropy to a localized profile, while keeping all other simulation parameters unchanged. Following predictions by Eq.~(\ref{2D_localized_anisotropy}), we initialize the bi-Maxwellian anisotropy of the 
form $T_{eyy}=T_{ezz}=10^{-4}m_ec^2$ and $T_{exx}=T_{eyy}+T_0\exp\left(-[(x-x_0)^2+(y-y_0)^2]/\delta^2\right)$, 
where $T_0=0.04m_ec^2$, $\delta=\sqrt{0.5}d_e$ and $(x_0,y_0)=(L_x/2,L_y/2)$ with the uniform density $n_0$. The cold ion temperatures are fixed 
uniformly at $10^{-4}m_ec^2$, the EM boundary conditions are Silver-M\"{u}ller, and particles are removed upon exiting the simulation boundaries 
to minimize boundary-induced artifacts. This form of localized anisotropy is especially relevant 
for scenarios involving localized phenomena such as laser-plasma interaction regions as encountered in recent studies~\cite{Zhao2024,Gregori2015}.

Fig.~\ref{2D_streak_xx_yy} shows streak plots of the various physical quantities from the 2D 
simulation. As in the above cases, all fields have been Gaussian filtered by 10 grid points. As a result, a quadrupole 
structure of the $B_z$ is clearly observed (Fig. \ref{2D_streak_xx_yy} (a)) as predicted by Eq. (\ref{2D_localized_anisotropy}). An associate quadrupole structure of $\Omega_{ez}$ is also generated, which combines with $B_z$ and yield $Q_{ez}$ (Figs. \ref{2D_streak_xx_yy} (b) and (c)). Further, Fig.~ \ref{2D_streak_xx_yy}(d) shows $\partial Q_{ez}/\partial t$, calculated using a backward difference scheme. Note that in this case, the data were recorded at sparse time intervals, and so the calculation of $\partial Q_{ez}/\partial t$ is relatively inaccurate. 
Fig. \ref{2D_streak_xx_yy}(e) shows the generation of $\mathcal{C}_z$ but is relatively weak as there is no initial drift in the system, confirming that the observed dynamics are solely driven by the pressure anisotropy term $\mathcal{B}_z$. Nevertheless, the strong similarity between Fig.~\ref{2D_streak_xx_yy}(d) and (f) further validates the proposed canonical vorticity model for analyzing such plasma dynamics. Specifically, it highlights 
that $\mathcal{B}_z$ in Eq.~(\ref{canonical_vorticity_eqn}) which reduces to Eq.~(\ref{canonical_battery_2Dz_local}), is 
the main source of the $\partial Q_{ez}/\partial t$. 

Further, the ${\mathcal{B}_{z}}$ in Eq.~(\ref{canonical_battery_2Dz_local}) consist of both ${p}_{exy}$ and $({p}_{exx}-{p}_{eyy})$ source terms, but the $2nd-$term is dominant, so it gives rise to a quadrupole 
structure of the growing filaments (Fig.~\ref{2D_streak_xx_yy} (a)-(c)). Both the source terms act together, giving a relatively higher growth rate of the instability, and are saturated by $t\omega_{pe}\simeq 10$. 
In contrast, when only the $1st-$term is active as in the case shown in Fig.~\ref{strek_plot_BWQ_pxy}, the instability develops slowly and saturates later around $t\omega_{pe}\simeq 100$. Furthermore, one observed flow characteristics is that $\mathcal{B}_z$ gives rise to localized quadrupole structures, which subsequently propagate along the $\pm \hat{x}$ direction. The sign and structure of the $\Omega_{ez}$ indicate an inflow along $\pm \hat{y}$ and outflow along $\pm \hat{x}$ directions, which are in direct correlation with the imposed anisotropic pressure distribution.
A detailed temporal analysis of the peak values of various quantities appearing in Eq.~(\ref{canonical_vorticity_eqn}) is discussed in our recent work~\cite{Laishram2024}.
\subsection{2D localized effects of non-uniform density and pressure anisotropy}
\label{nonuniform_sim2D}
\begin{figure*}
\includegraphics[width=\textwidth]{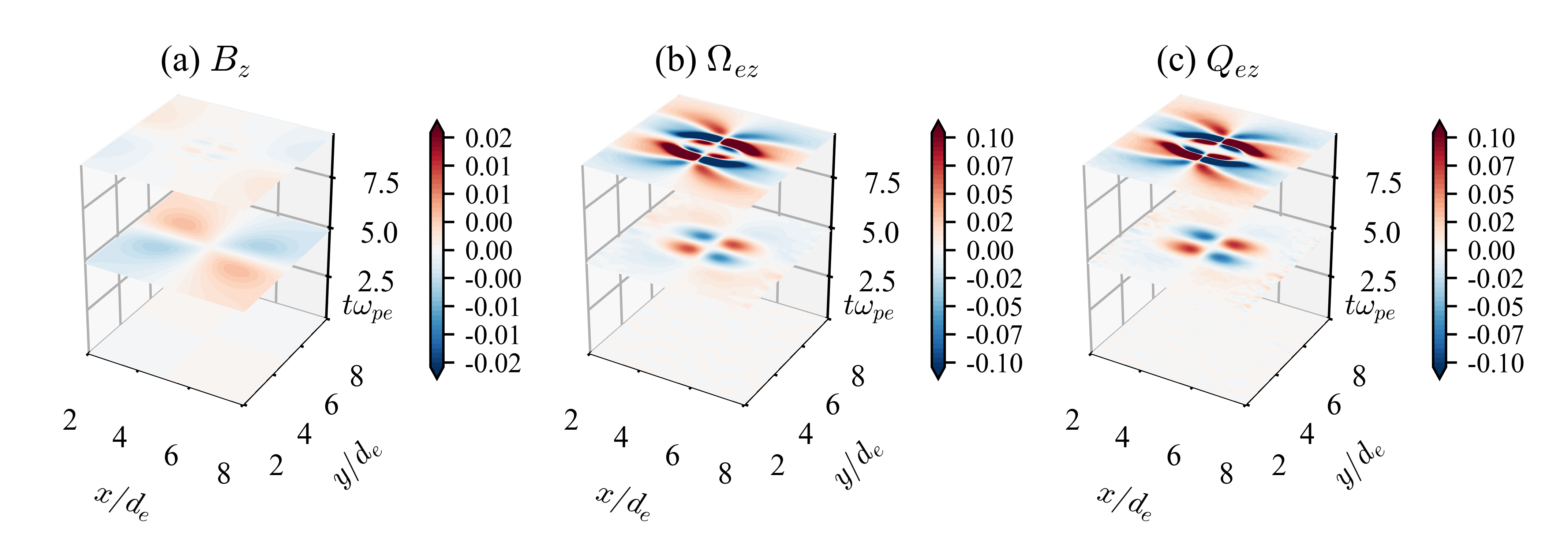}\\
\caption{Streak plots of  (a) $B_z$, (b) $\Omega_{ez}$, and (c) $Q_{ez}$ from the 2D PIC simulation for coexistence of localized density $n_e=n_0 \exp\left(-[(x-x_0)^2+(y-y_0)^2]/2\delta^2\right)$ and 
non-localized temperature anisotropy $T_{eyy}=T_{ezz}=10^{-4}m_ec^2$ and $T_{exx}=0.04 m_ec^2 $. }
\label{2D_streak_xx_yy_dn_Tyz}
\end{figure*}

\begin{figure*}
\includegraphics[width=\textwidth]{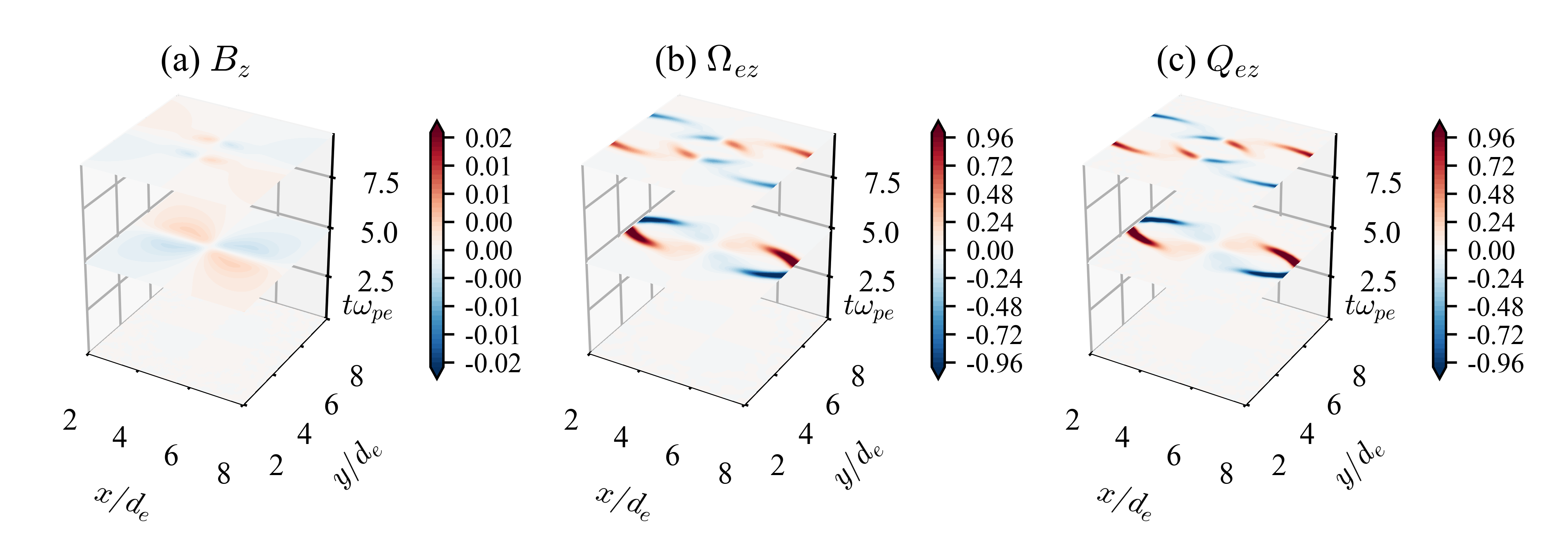}\\
\caption{Streak plots of  (a) $B_z$, (b) $\Omega_{ez}$, and (c) $Q_{ez}$ 
from the 2D PIC simulation of coexistence of localized density $n_e=n_0 \exp\left(-[(x-x_0)^2+(y-y_0)^2]/2\delta^2\right)$ and 
localized temperature anisotropy $T_{eyy}=T_{ezz}=10^{-4}m_ec^2$ and  
$T_{exx}=T_{eyy}+0.04m_ec^2\exp\left(-[(x-x_0)^2+(y-y_0)^2]/\delta^2\right)$. }
\label{2D_streak_xx_yy_dn}
\end{figure*}

In addition to the setups presented above, a whole menagerie of pressure-tensor configurations that lead to magnetogenesis can be generally predicted using the canonical battery terms. For example, initially if a strong density inhomogeneity $n_e=n_0 \exp\left(-[(x-x_0)^2+(y-y_0)^2]/2\delta^2\right)$ and the non-localized electron temperatures anisotropy of $T_{eyy}=T_{ezz}=10^{-4}m_ec^2$ and $T_{exx}=0.04 m_ec^2 $ coexist, additional effect of the $3rd-$term of Eq.~(\ref{canonical_battery_2Dz}) become significant. In such case the additional dominant term, assuming $p_{eij} = n_e T_{eij}$, is
\begin{equation}
    \mathcal{B}_z \sim T_{exx}\left(\frac{\partial}{\partial y}  \frac{1}{n_e}\right) \left( \frac{\partial {n}_{e} }{\partial x}\right),
    \label{reduc_1}
\end{equation}
which has the dependence $\propto -xy$ and so it must 
oppose the pressure anisotropy effect of Eq.~(\ref{canonical_battery_2Dz_local}). Fig.~\ref{2D_streak_xx_yy_dn_Tyz} shows streak plots of $B_z, ~\Omega_z$ and $Q_{ez}$ from the corresponding 2D simulation. As predicted, a reduction in the magnitude of both the growing $B_z$ 
and $\Omega_z$ quadrupole structure up 
to $t\omega_{pe}\approx 5$ is observed compared to above case in Fig.~\ref{2D_streak_xx_yy}. An additional effect of spreading the structures uniformly in radial directions is also observed due to the diffusion of the initially localized density. Thus $B_z$ evolves smoothly with larger scale, while $Q_{ez}$ (due to $\Omega_z$) is tied to stronger gradients and so more complex features with filamentary periodic patterns of 
spreading outward from the localized region.

In another case, initially if the strong density inhomogeneity $n_e=n_0 \exp\left(-[(x-x_0)^2+(y-y_0)^2]/2\delta^2\right)$ co-exist with the localized temperature anisotropy considered in Fig.~\ref{2D_streak_xx_yy}, the additional effect of 
the $3rd-$term of Eq.~(\ref{canonical_battery_2Dz}) become, 
\begin{equation}
     \mathcal{B}_z \sim \left(\frac{\partial}{\partial y}  \frac{1}{n_e}\right) \left( \frac{\partial {p}_{exx} }{\partial x}\right),
    \label{reduc_2}
\end{equation}
which also has the dependence $\propto -xy$ and so opposes the original source Eq.~(\ref{canonical_battery_2Dz_local}). Here, Eq.~(\ref{reduc_2}) involves gradients in both $n_e$ and $T_{exx}$, therefore its magnitude should be stronger than the above case of Eq.~(\ref{reduc_1}), which means that this case will have more reduction effects. Fig.~\ref{2D_streak_xx_yy_dn} shows streak plots of $B_z, ~\Omega_z$ and $Q_{ez}$ from its corresponding 2D simulation. As predicted, a strong reduction in the magnitude of both the growing $B_z$ and $\Omega_z$ quadrupole structure up to $t\omega_{pe}\approx 5$ is observed compared to the case in Fig.~\ref{2D_streak_xx_yy} and Fig.~\ref{2D_streak_xx_yy_dn_Tyz}. While $B_z$ is relatively reduced, $\Omega_z$ turns out stronger by one order, and the diffusion effect is non-uniform with elongated bipolar lobes along $\hat x$ as in the 
case of Fig.~\ref{2D_streak_xx_yy}. Similarly, there can be multiple structures of ${Q}_{ez}$ from all possible combinations of the source terms in Eq.~(\ref{canonical_battery_2Dz}).
\begin{figure*}
\includegraphics[width=1.5\columnwidth]{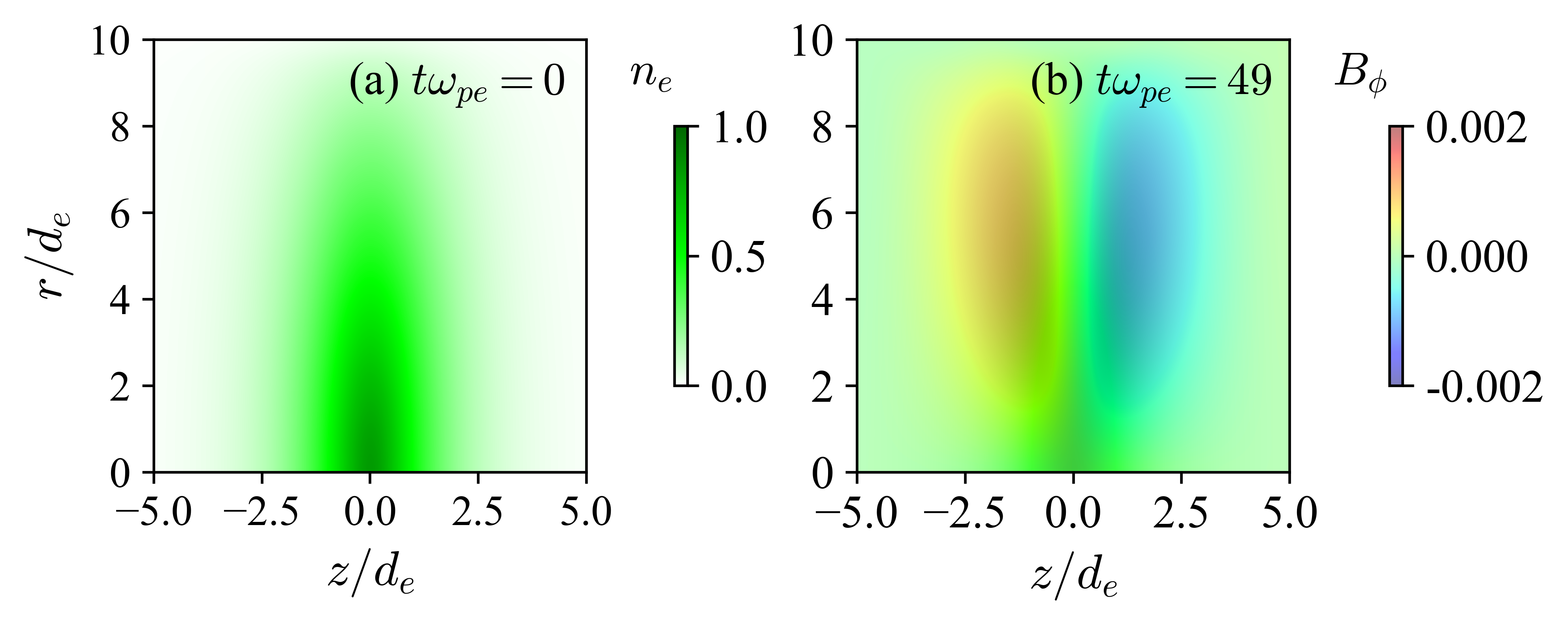}\\
\caption{Overlaying slice plots of density $n_e [\epsilon_0 m_e\omega_{pe}^2/e^2]$ on the top of 
$B_\phi [m_e\omega_{pe}/e] $ at $(a) ~t\omega_{pe}=0$ and 
$(b) ~t\omega_{pe}=49$ from the 2D PIC simulation in axisymmetric cylindrical geometry.}
\label{Beirman_mag}
\end{figure*}

\subsection{2D kinetic scale Biermann battery effect}
\label{Beirman_sim2D}
We have explored various tensorial pressure anisotropy effects arising from non-uniform density and temperature anisotropy. Now, we can see a kinetic form of the Biermann battery effect in a plasma, driven solely by its baroclinic nature (with temperature isotropic), as predicted in Eq.~(\ref{Biermann_eqn2}).
To investigate this, the above simulation is extended to an axisymmetric cylindrical geometry with domain size $(L_r, L_z)=(10,10)d_e$, retaining the same grid sizes and time-steps as before. Then consider an initial isotropic temperature has radial variation of the form of natural Bessel mode $T=T_0J_0(\alpha_0r/L_r)$, where $T_{0}=0.04 m_ec^2$, $\alpha_0=2.4048$ (first zero of $J_0$). The density is initialized with a 2D Gaussian profile of the form $n_e=n_0\exp\left(-[(r/\delta_r)^2+(z/\delta_z)^2]\right)$, 
where $n_0=1$, $\delta_r=5d_e$, and $\delta_z=1d_e$ are the radial and axial characteristic widths of the shear profile.
This setup is relevant to many of the laser-plasma experiments, where a non-vanishing baroclinic term $\nabla{n_e }\times\nabla {p_e}\ne0$ naturally arises at the focal region of intense laser pulses-substrate interaction regions~\cite{Gregori2015,Matteucci2018,Jinno2022,Zhao2024}. The ion temperatures are set 
uniformly to $10^{-4}m_ec^2$ as in previous cases, and boundary conditions are set to ``remove" for particles 
and ``open" for electromagnetic fields.

Fig.~\ref{Beirman_mag} shows slice plots of both the 
electron density overlaying on the outcome azimuthal magnetic field ${\bf B}_\phi$ at two different timesteps along the development of the Barmann battery effect. Initially, ${\bf B}_\phi$ is zero, and electrons are accumulated around the origin as shown in Fig.~\ref{Beirman_mag}(a). 
Later, as described by Eq.~(\ref{Biermann_eqn2}), electron density is gradually diffused out and ${\bf B}_\phi$ emerges as the system evolves due to $\nabla{n_e} \times  \nabla {p_e}\ne0$ as shown in Fig.~\ref{Beirman_mag}(b). Further, it is observed that the spatial structure and peak location of ${\bf B}_\phi$ are strongly influenced by the relative radial profiles of density and temperature with respect to the system size. The generated ${\bf B}_\phi$ is relatively weak, smooth, and unidirectional in the whole $(L_r, L_z)$ domain, unlike the above Weibel instability cases where intense filamentary and quadrupole patterns dominate, highlighting the fundamentally different characteristics of the two distinct 
processes. A more detailed analysis of ${\bf B}_\phi$-generation in terms of canonical vorticity will require a cylindrical co-ordinate formulation of Eq.~(\ref{canonical_battery_2Dz}) and is deferred for future study.

\begin{figure*}
\includegraphics[width=\textwidth]{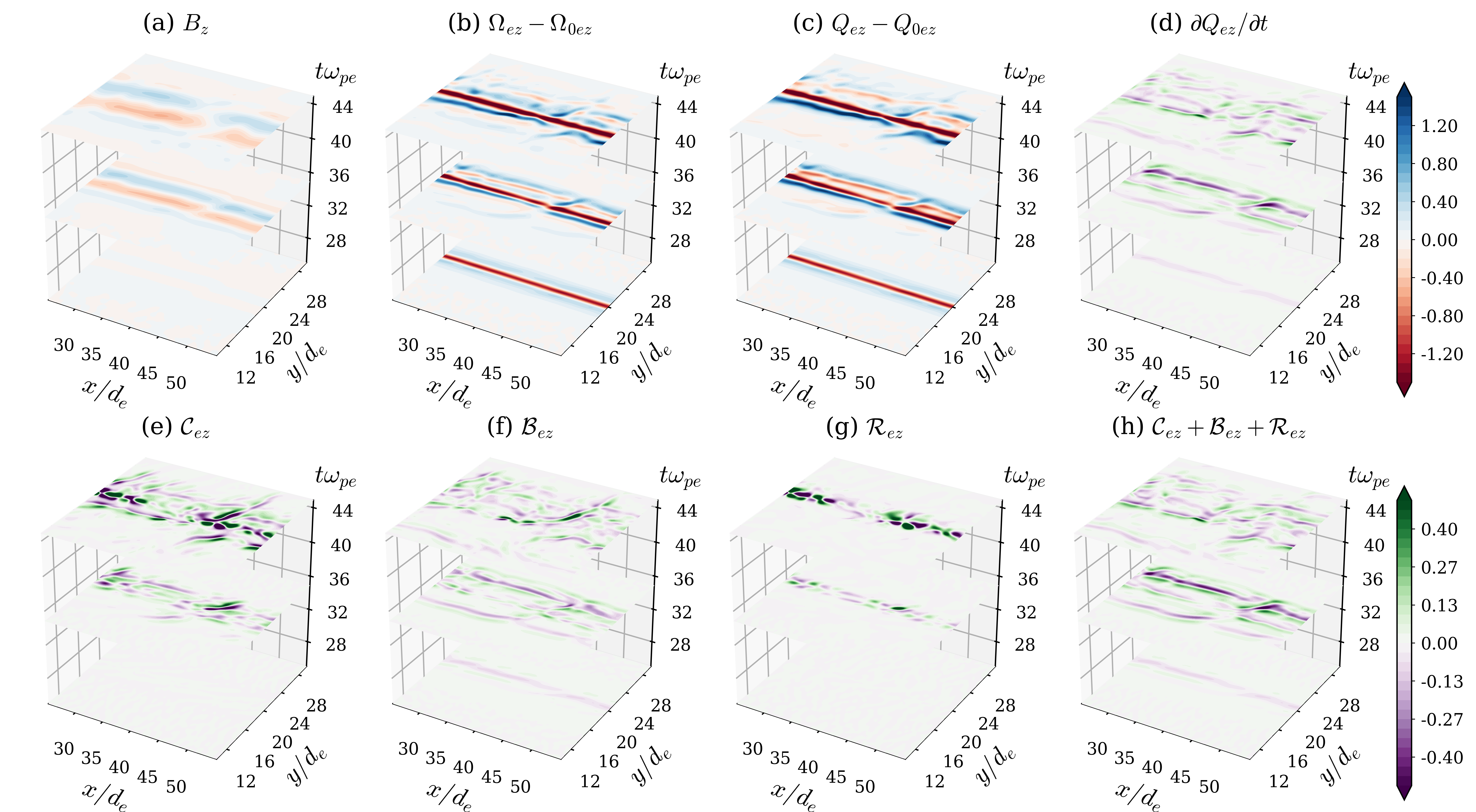}\\
\caption{Streak plots of various quantities 
(a) $B_z[m_e\omega_{pe}/e]$, 
(b) $(\Omega_{ez}-\Omega_{0ez})[m_e\omega_{pe}]$, 
(c) $(Q_{ez}-Q_{0ez})[m_e\omega_{pe}]$, 
(d) $\partial Q_{ez}/\partial t [m_e\omega_{pe}^2]$,
(e) $\mathcal{C}_z~[m_e\omega_{pe}^2]$,  
(f) $\mathcal{B}_z~[m_e\omega_{pe}^2]$, 
(g) $\mathcal{R}_{ez} ~[m_e\omega_{pe}^2]$, and 
(h) ($\mathcal{C}_{ez}+\mathcal{B}_{ez}+\mathcal{R}_{ez})~[m_e\omega_{pe}^2]$ from the 2D PIC simulation. The three slices in each panel correspond to $t\omega_{pe}=25,35,$ and $45$ (reproduce from~\citet{Laishram2025}).}
\label{2D_streak}
\end{figure*}

\subsection{2D view of relativistic kineclinicity effect}
\label{relativistic_sim2D}

We have shown in Eq.~(\ref{canonical_vorticity_eqn_rel}) that relativistic effects add new source/sink term ${\vec{\mathcal{R}}}$ for the ${\bf Q}_\sigma$, in additional to the canonical battery (${\vec{\mathcal{B}}}$)-effects. 

To validate the model and visualize the role of each term in Eq.~(\ref{canonical_vorticity_eqn}), several 2D3V particle-in-cell (PIC) simulations were conducted for a modified relativistic Beltrami flow of a pair plasma in a domain $(L_x, L_y)=(80,80)~d_e$. The initial fluid velocity profile for both electrons and positrons were
$u_x=-u_{x0}[\tanh\left(\left\{y-0.25L_y\right\}/\delta\right)-\tanh\left(\left\{y-0.75L_y\right\}/\delta\right)-1], 
u_y=0$,
and $u_z=u_{z0}[\textrm{sech}\left(\left\{y-0.25L_y\right\}/\delta\right)-\textrm{sech}\left(\left\{y-0.75L_y\right\}/\delta\right)]$,
with $\delta=1d_e$, a uniform initial density $n_0$ and thermal velocity $v_{th}/c=0.097$, and periodic boundary conditions were used. The details of the simulation and comprehensive study of the role of ${\vec{\mathcal{R}}}$ are discussed in our recent work~\cite{Laishram2025}. 

For $u_{x0}=u_{z0}$, this flow satisfies the Beltrami flow condition $\nabla\times\mathbf{u}=\lambda\mathbf{u}$, where $\lambda$ is a scalar function \cite{Mahajan1998}, and susceptible to both ESKHI and Weibel instability, depending on the underlying parameters. However, we choose initial conditions for which all the terms in Eq.~(\ref{canonical_vorticity_eqn_rel}) are roughly equally important. To satisfy this condition, the fiducial run was empirically chosen to have $v_{z0}=c\sqrt{1-\gamma^{-2}}$ where $\gamma=50$, and $v_{x0}=v_{z0}/1.5$. This condition may be relevant to relativistic jet along the $z$-direction with a finite hydrodynamical helicity density~\cite{Liang_2013, Li2006, Bonafede2010}.

Then, for detailed visualization, we focus on the electron dynamics at a particular shear region at $y=0.25L_y$, which is similar to the mirror reflection of the shear region at $y=0.75L_y$. Fig.~\ref{2D_streak} shows streak plots of various quantities from the fiducial simulation. As usual, all quantities have been Gaussian filtered by 10 grid points to reduce the noise arising from taking multiple gradients. A current-sheet like $B_z$ structure (Fig.~\ref{2D_streak}(a)) is self-generated along with a relatively strong $\Omega_{ez}$  (Fig.~\ref{2D_streak}(b)), which together yield $Q_{ez}$ (Fig.~\ref{2D_streak}(c)). Fig.~\ref{2D_streak}(d) shows $\partial Q_{ez}/\partial t$, calculated with a backward difference scheme. This corresponds to the left-hand-side of Eq.~(\ref{canonical_vorticity_eqn_rel}).

Figs.~\ref{2D_streak}(e-h) show the streak plot of various terms on the right-hand side of Eq.(\ref{canonical_vorticity_eqn_rel}) and their total sum. The strong resemblance between Fig.~\ref{2D_streak} (d) and (h) validates the canonical vorticity formulation given in Eq.~(\ref{canonical_vorticity_eqn_rel}). In contrast with the above non-relativistic cases, $\mathcal{C}_{ez}$ in Fig.~\ref{2D_streak}(e) turns out strong complex structure due to the initial perturbed Beltrami flow condition. The strong variation between Fig.~\ref{2D_streak}(e) and Fig.~\ref{2D_streak}(h) indicates that the frozen-in property of $Q_{ez}$ is significantly violated during the evolution. Comparing Fig.~\ref{2D_streak}(e) to Fig.~\ref{2D_streak}(g), $\mathcal{R}_{ez}$ roughly cancels out $\mathcal{C}_{ez}$ around the highly shear region, showing a relativistic suppression role against the ${\mathcal{ C}_{\sigma}}$-effects, whereas $\mathcal{B}_{ez}$ is relatively weak within this region. 

We have shown that different physical mechanisms generate seed magnetic fields spontaneously across distinct spatial and temporal scales, potentially influencing subsequent dynamo amplification.
The magnetic field strength observed in Fig.~\ref{Spatiotemporal_all_By_case}~to~\ref{2D_streak_xx_yy_dn}, $B_z \approx 0.02~[m_e\omega_{pe}/q_e]$, agrees with values typical of Weibel-instability-driven fields reported in laser-plasma experiments~\cite{Zhao2024, Gregori2015,Huntington2015}. For example, in dense plasmas with $n_0 = 10^{19}$ cm$^{-3}$ representative of laser-solid interactions~\cite{Huntington2015}, this corresponds to $\sim$20 T, matching experimental observations and supporting Weibel-driven magnetogenesis under such conditions.
Likewise, the self-generated field in Fig.~\ref{2D_streak}, $B_z \approx 0.5~[m_e\omega_{pe}/q_e]$, is comparable to that produced in relativistic shear flows of hybrid $e^+e$–ion plasmas~\cite{Liang_2013}. For similar relativistic shear flows with $n_e \approx 10^{10}$ cm$^{-3}$, relevant to GRBs, this yields seed fields of $B_z \approx 10^2$-$10^3$ G consistent with magnetic-field estimates for GRBs, the intracluster medium, and cosmological environments~\cite{Liang_2013, Li2006, Takahashi2005, Bonafede2010, Vachaspati2021}.

\section{Discussions}
\label{Discussions}
One of the key strengths of the 2D-canonical vorticity model Eq.~(\ref{canonical_vorticity_eqn}) or Eq.~(\ref{canonical_vorticity_eqn_rel}) is its compact structure derived without simplifying assumptions; each term on the right side corresponds to a distinct kinetic process. For each species, the term $\mathcal{C}_{\sigma z}$ describes inertial flux-conserved convection flow or 
dynamo-driven amplification process, while the 
terms $\mathcal{B}_{\sigma z}$ and $\mathcal{R}_{\sigma z}$ represent the seed-field generation processes. A simple scaling comparison of $\mathcal{C}_{\sigma z}$ and $\mathcal{B}_{\sigma z}$ terms allows us to estimate the length scale at which the transition occurs from generation to amplification process. For example, for $e-$species
with cold-ion plasma, assuming the $\mathcal{C}_{\sigma z}$ term becomes the 
dynamo-effect when $|\mathbf{u}_e| \sim v_A$ (Alfv\'en velocity) and 
$\mathbf{Q}_e \sim q_e \mathbf{B}$ as follows,

\begin{align}  \nonumber
    |\mathbf{u}_e\times \mathbf{Q}_e|: \nabla\mathcal{P}_e/n_e,\\  \nonumber
    |{v}_A\times q_e\mathbf{B}_e|: L^{-1} \mathcal{P}_e/n_e,\\ \nonumber
    |(\mathbf{B}_e/\sqrt{\mu_0n_im_i})\times q_e\mathbf{B}_e|: L^{-1} \mathcal{P}_e/n_e,\\
    \Rightarrow L \approx \mu_0\mathcal{P}_e/\mathbf{B}_e^2 \sqrt{m_i/\mu_0n_iq_e^2}\approx \beta_e d_i/2 .
     \label{scaling_WB_dynamo}
\end{align}
Where, $\beta_e=2\mu_0\mathcal{P}_e/B^2$ is the electron plasma beta and $d_i=\sqrt{m_i/\mu_0n e^2}$ is ion skin depth. It 
shows that the seed-field generation ($\mathcal{B}_{ez}$-effect) is dominant over amplification ($\mathcal{C}_{ez}$-effect) for characteristic lengths up 
to $L\approx \beta_e d_i/2$. Thus, by examining the interplay between the two terms, one can identify the progression of magnetic fields from 
micro to macro scales. Such transitions have been gaining recent attention~\cite{Sironi2023,Zhou2024} and are worth exploring in the canonical vorticity framework, especially because the merging or breaking of Weibel-induced filaments involves reconnection, which is controlled by both convective and battery terms~\cite{Yoon2017,Yoon2018,Yoon2019a}. 

Similarly, $\mathcal{B}_{ez}$ (in Eq.~(\ref{canonical_battery_2Dz})) itself has three different sub-terms that induce different phenomena, namely Weibel instability, Biermann battery effect, and localized pressure anisotropy effect~\cite{Laishram2024}. Their relative comparison reveals primary and secondary mechanisms evolve and their transitions along the plasma dynamics. {\citet{Schoeffler2014}} showed that the relative importance of the Biermann effect to the Weibel instability depends on the length-scale of the source. Such a transition around a system length scale $L$ can simply be explained by the ratio of the related sub-terms in $\mathcal{B}_{ez}$ as follows,
\begin{align}  \nonumber
    \left|\frac{\nabla n_e\times \nabla T_e}{n_e}\right| &: \frac{1}{n_e}\frac{\partial^2 \mathcal{P}_{exy}}{\partial x^2},\\
    \frac{T_{diag}}{T_{off}}d_e^2 &: L^2 ,
     \label{scaling_WB}
\end{align}
where we have assumed $\nabla^{-1} \sim L\sim L_n \sim L_T$, and $T_{diag}$ and $T_{off}$ are diagonal and off-diagonal components of the temperature tensor, respectively. The Weibel instability arises at $d_e$-scales, so $\partial^2 \mathcal{P}_{exy}/\partial x^2\sim \mathcal{P}_{exy}/d_e^2$. It can be figured out that $T_{diag}$ is around one or two orders of magnitude bigger than $T_{off}$, and so the transition is at around $L\sim 10 ~d_e$~\cite{Laishram2024}. Such a comparison explains the scale-dependent dominance of seed-field generation mechanisms.

The extension of the model in canonical momentum space (${\bf p}$-field) in Eq.~(\ref{canonical_vorticity_eqn_rel}) generalizes the idea of magnetogenesis by adding a new source, i.e., $\mathcal{R}_{ez}$-effect to already established $\mathcal{B}_{ez}$-effects~\cite{Laishram2024}. Although $\mathcal{R}_{ez}$ partially cancels 
out $\mathcal{C}_{ez}$ in our recent work~\cite{Laishram2025}, this does not imply that kineclinicity always suppresses the canonical vorticity. For instance, if $\mathcal{C}_{ez}$ is cancel out with $\mathcal{B}_{ez}$ in 
Eq.~(\ref{canonical_vorticity_eqn_rel}), then $\mathcal{R}_{ez}$ can be big. This means that there may be situations where kineclinicity is dominantly responsible for canonical vorticity generation. We have been trying to isolate such a configuration, but to no avail so far, so it remains to be seen whether a kineclinicity-dominated magnetogenesis can occur. Even a relatively weak independent $\mathcal{R}{ez}$ term could still generate cosmological seed fields in relativistic shear flows~\cite{Takahashi2005}. Moreover, $\mathcal{R}_{ez}$-effects may be 
an important fundamental feature of many relativistic scenarios, including astrophysical jets, collisionless shocks, relativistic magnetic reconnection, and many more. In particular, our preliminary study shows that the effect may be significant in relativistic magnetic reconnection, where the kineclinic term may be an additional source of topology change of $\mathbf{Q}$ and thus $\mathbf{B}$ field lines. A detailed understanding of its roles would require further extensive studies. 

A limitation of the current formulation is that Eq.~(\ref{canonical_vorticity_eqn_rel}) is not yet expressed in a 
covariant form~\cite{Mahajan2010}. Since the underlying physics-governed by the Vlasov equation and Lorentz force, can be formulated in a fully covariant form, it would be interesting to cast this formulation in a covariant form. Such a covariant reformulation would clarify the role of $\mathcal{R}_{\sigma z}$ in curved space-time and ensure consistency with general-relativistic plasma theory.

Further, the relativistic simulation focuses on pair plasmas to avoid the multiscale complexity and computational stiffness associated with the $m_i/m_e$ ratio. In electron-ion plasmas, one would expect substantially richer structure formation and additional instabilities~\cite{ Grismayer2013, Nishikawa2014}. Since Eq.~(\ref{canonical_vorticity_eqn}) applies to any species, an ion-driven kineclinic effect should arise analogously but on longer timescales. Recent work indeed shows ion contributions emerging after electron-induced fields saturate~\cite{Yoon2025PoP}.

All the above simulations were performed in 2D setups with optimized parameters, and observed qualitatively similar results as in most of the 
relevant studies in 2D and 3D systems~\cite{Matteucci2018, Liang2013, Nishikawa2014,Zhou2024,Romanov2004,Schoeffler2014}. However, realistic astrophysical magnetogenesis requires full 3D analysis, which we plan for future work.
Moreover, the present work so far focuses only on the linear growth phase of magnetic field generation; future studies should explore the nonlinear saturation mechanisms and the onset of turbulence in both 2D and 3D systems. Such investigations will be crucial for understanding magnetic field evolution in both laboratory experiments and astrophysical plasmas~\cite{Achterberg2007, Schekochihin2009, Grassi2017, Zhou2024}.

\section{Summary}
\label{summary}
We briefly review the current status of magnetogenesis, a cross-disciplinary research field at the interface of cosmology and plasma physics, which investigates the origin and amplification of magnetic fields in plasmas. We place particular emphasis on plasma-physics processes and formulate a canonical vorticity framework for collisionless plasmas to study magnetogenesis at the kinetic level. By treating canonical vorticity as the primary dynamical variable, this approach unifies and generalizes several magnetogenesis mechanisms, such as the Biermann battery effect and the Weibel instability, and further predicts more complex pressure-tensor configurations, altogether as a different form of canonical battery effect that acts as a fundamental source of the canonical vorticity, which is partly magnetic field
The framework is further extended to the relativistic regime and reveals a new source of the canonical vorticity arising from the misalignment between fluid momentum and velocity gradients, referred to as the kineclinicity effect. This effect has no non-relativistic counterpart and may be an important fundamental feature of many relativistic scenarios, and therefore, this formulation calls for further investigation into the role of kineclinicity in a broad range of relativistic environments.
Several of the predictions by the framework, both in non-relativistic and relativistic regimes, are systematically verified through a series of PIC simulations, and discussed several implications of the framework and its limitations. Shortly, this framework will apply to studies of more extreme plasma characteristics relevant to high-energy laboratory experiments and astrophysical plasma environments.

\begin{acknowledgments}
This work was supported by an appointment to the JRG Program at the APCTP through the Science and Technology Promotion Fund and Lottery Fund of the Korean government, and also by the Korean Local Governments---Gyeongsangbuk-do Province and Pohang City. This work was also supported by the NRF of Korea under Grant No. NRF-RS-2023-00281272 and NRF-RS-2025-00522068. The computations presented here were conducted on the KAIROS supercomputing cluster at the Korean Institute of Fusion Energy and on the APCTP computing server.
\end{acknowledgments}

\bibliography{export2}
\end{document}